\documentclass{siamltex}
\usepackage{amssymb}
\usepackage{amsmath}
\usepackage{graphicx}
\usepackage[square]{natbib}

\def\R{\mathbb{R}}
\def\Rp{\mathbb{R^+}}

\newtheorem{defn}[theorem]{Definition} %uncomment for siamltex.cls and see end of abstract

\renewcommand{\thetable}{\Roman{table}}

\begin{document}

\title{ Optimal intervention in the foreign exchange market when interventions
affect market dynamics }

\author{
	Alec N. Kercheval\footnote{Corresponding author. Address: Department of Mathematics, 1017 Academic Way, Room 208,
	Florida State University, Tallahassee, FL 32306-4510. email: {\tt kercheva@math.fsu.edu}, phone:
	850-644-8701, fax: 850-644-4053.} 
\and
	Juan F. Moreno\footnote{Address: Calle 103 A \# 11 B 49 Bq1 Apt 302, Bogota, Colombia.
	email: {\tt jmoreno@math.fsu.edu}, phone: 57 1 6370221.} 
	}
		
\date{March 13, 2009}

\maketitle

Abbreviated Title: Optimal intervention in foreign exchange

\begin{abstract}
We address the problem of optimal Central Bank intervention in the exchange rate market when interventions
create feedback in the rate dynamics. In particular, we extend the work done on optimal impulse control by Cadenillas
and Zapatero \cite{cadenillas-zapatero99, cadenillas-zapatero00}  to incorporate temporary market reactions, of random duration and level,
to Bank interventions, and to establish results for more general rate processes.   We obtain new explicit optimal impulse control strategies that account for these market reactions,
  and show that they cannot be obtained simply by adjusting the intervention cost in a model
  without market reactions.
  \end{abstract}
 %  {\em AMS 2000 Mathematics Subject Classifications:}  93E20, 91B70, 60G40.


\medskip

\section{Introduction}

In countries dependent on foreign trade and foreign capital, the Central Bank is normally in charge of
exchange rate policy.  This usually means that the Central Bank has the ability to intervene in
the markets in order to keep their currency rates within a band, or close to a target rate set
by the Bank's policy makers.

Intervention can take two different (compatible) forms: adjustment of domestic interest rate levels, which
influences the attractiveness of foreign investments; and direct purchases or sales of foreign currency
reserves in the foreign exchange market.

The first form of intervention can be modeled as a continuous (classical) control problem, and
the second as an impulse control problem.
In this paper we focus only on the second type of intervention, interpreting a market intervention as
a way to change the exchange rate by a desired amount via sales or purchases of reserves
over a time short enough to be reasonably modeled as an instantaneous impulse.

The problem is to find an optimal intervention strategy keeping the exchange rate near a target level,
set by the bank, while minimizing cost of intervention.  Since the exchange rate will always
drift away from the target rate between interventions, one
approach to control intervention costs has been to set
a ``target zone" or band, and act to keep the exchange rate within this band. Various papers in the
economics literature have considered this problem, e.g. Krugman \cite{krugman91}, Froot and Obstfeld \cite{froot-obstfeld91},
and Flood and Garber \cite{flood-garber91}.

The first person to apply the theory of stochastic impulse control to this problem was Jeanblanc-Picque
\cite{jeanblanc-picque93}, later extended by Korn \cite{korn97}.  Both considered an exogenously
specified target band within which the exchange rate is to be contained, and found the optimal
sizes of interventions required when the exchange rate reaches the boundary of the target band.

An important insight was obtained by Mundaca and Oksendal \cite{mundaca-oksendal98} and Cadenillas and Zapatero \cite{cadenillas-zapatero99, cadenillas-zapatero00},
who realized that it is not necessary to exogenously set a target band. Rather, the
correct target band can be derived endogenously as part of the solution to the optimization
problem, using a cost function combining cost of intervention with
a running cost given by an increasing function of the distance between the target rate
and the current rate. Mundaca and Oksendal used a standard brownian motion model for the 
underlying exchange rate;  Cadenillas and Zapatero used a geometric brownian motion and explicitly computed, for certain examples, the optimal intervention
levels and intervention amounts.

They expressly assume, however, that investors do not observe or anticipate
the interventions of the Bank, so that {\em the process driving the rate dynamics is not affected by
interventions.}  This is unrealistic, but to do otherwise ``would yield different dynamics for the exchange rate and would probably
make the model intractable'' \citep{cadenillas-zapatero00}.

  In this paper we overcome
this intractability to solve the problem with
the same level of explicitness as do Cadenillas and Zapatero, but for more general
exchange rate processes, and -- most importantly --
allowing for a market reaction
to interventions.  We assume that the rate dynamics changes to a different process for
a random period of time $T$ after each intervention (where $T$ is assumed independent
of the rate process), after which it reverts to the pre-intervention process.  For example,
we could imagine that the volatility of the exchange rate might move to a new, higher level
for a period of time after an intervention, to reflect heightened market uncertainty about
the path of rates in the near term. The new rate process need not be known in advance: it can be
drawn at random each time an intervention takes place, as long as the draw is $iid$ and
also independent of the rate process.

We still must retain the assumption that investors do not anticipate interventions by the Bank.
This is much milder than the assumption that investors do not {\em react} to interventions.  Indeed,
we might expect that the Bank itself will be revising the parameters used in it's exchange rate model
as time passes, so that the optimal solution today would not persist due to the incorporation
of new information in the Bank's model.
  If this happens, investors are unlikely to
have much confidence in forecasts of the next Bank intervention.  Hence it is not necessary
to assume that interventions are invisible to the market in order to reasonably ignore the market
effect of investor prediction of future intervention times.

 %%%

 We solve our problem by applying the theory of stochastic impulse control (see Korn \cite{korn99} for
 a good overview) and, as we are able to obtain analytical results, we can compare the policies with and without a reaction period. In the spirit of Cadenillas and Zapatero, and since we can provide numerical solutions, we also include comparative statics analysis about the effects of the changes of parameters on the optimal intervention strategy.

We consider a currency with exchange rate dynamics modeled by a general It\^o diffusion, such
as a geometric Brownian Motion, which temporarily changes, at random, to a different It\^o diffusion
during a ``reaction period" that lasts for a finite random time
$T$ after each intervention by the Central Bank. The Bank tries to keep this exchange rate close to a given target, and there is a running cost associated to the difference between the exchange rate and the target. However, there are also fixed and, optionally, proportional costs associated with each intervention. 
The Bank determines when and by how much to move the rate, but cannot control the duration
of the reaction period or the rate process during the reaction period, neither of which are assumed
known by the Bank in advance.
The objective of this paper is to find the optimal level of intervention, as well as the optimal sizes of the interventions, so as to minimize the total cost.

Our analysis currently requires us to impose the restriction that the Bank is not allowed to intervene during the temporary reaction period. However, this restriction is reasonable from the perspective
of Central Bank policy. The reaction periods are intended to model short market
re-adjustment periods, so the restriction is short-lived. Also, after they have reset the rate to
a desired target, Central Banks will want to wait for a time to observe the medium-term effects
of their action.  If they think the market is still in a temporary reaction mode with uncertain
parameters, but will soon revert to the previous dynamic (volatility, drift), it is reasonable for
them to preserve their capital and wait for the re-establishment of the long term dynamical rate
parameters before contemplating another intervention.

We shall prove that when the exchange rate lies in a specific interval - the continuation region - the Central Bank's optimal policy is not to  intervene. However, when the exchange rate reaches the boundary of that interval -- when it enters the intervention region-- then the Central Bank must intervene (as soon as the reaction
time has expired), pushing the exchange rate to yet another interval, the preferred region inside the continuation region. (In case there are no
proportional intervention costs, the preferred interval degenerates to a single point.)

We illustrate an example in which a temporary reaction period of increased volatility leads to an optimal
policy in which the intervention band is widened, calling for great patience by the Central Bank, and less
frequent but larger interventions.  We show with an example that the optimal policy cannot be
calculated simply by increasing the intervention cost -- an increased cost can match the wider no-intervention
band, but will not reproduce the correct optimal restarting value. {\em Therefore, correct
intervention policy requires the modeling of market reactions to intervention.}

Korn \cite{korn97} also studies the case where interventions have random consequences, but in his model
it is the amount by which the rate is changed that is random, whereas here we assume the Bank
can control the initial rate change level, but cannot predict the new volatility or drift, or the duration
of the reaction period.

The structure of the paper is as follows: in section 2 we formulate the problem by introducing the exchange rate dynamics and the Central Bank objective; in section 3  we state the quasi-variational inequalities for this problem and state sufficient conditions of optimality for this impulse control problem; in section 4 we solve the problem of Central Bank intervention, with and without recovery period, and we perform some comparative statics analysis.  In section 5 we present the proof of the main theorem, and we close the paper with some conclusions.

	\section{Problem Statement}

We denote by $X_y(t)$ the exchange rate process in the absence of interventions, representing domestic currency units per unit of foreign currency at time $t$, with initial value
$X_{y}(0-) = y$. We suppose that $X_{y}$ follows the diffusion process given by the time-homogeneous stochastic differential equation
	\begin{equation}\label{eqn:GBM}
			dX_{y}(t) = \mu_1 (X_{y}(t)) dt + \sigma_1 (X_{y}(t)) dW_t,
	\end{equation}
where $W_t$ is a one-dimensional Brownian motion in a probability space $(\Omega, \mathcal{F},P)$ with augmented
natural filtration $\{ \mathcal{F}_t \}$, and $\mu_1, \sigma_1$ are Lipschitz functions on $\R$. 
	(If $\mu_1(x)>0$ the currency experiences a devaluatory pressure and if $\mu_1(x)<0$ a pressure to appreciate.)
	For technical reasons
mentioned below we suppose that the process $X_y$ is defined on a time interval $(-a, \infty)$, for $a>0$. 
The interval $[0, \infty)$ will be the domain of permitted intervention times.
%As a typical example for numerical illustration, we will look at the
%special case of geometric brownian motion when $\mu_1$ and
%$\sigma_1$ are linear, but our results apply to this more
%general setting.

Suppose now that interventions take place at times $\tau_i$, $i \in N = \{1,2,3,\dots \}$, such that
$$0 \leq \tau_1 < \tau_2 < \tau_3 < \cdots, $$  each
such intervention causing a discontinuous change in the exchange rate, where the intervention times and discontinuity
sizes are under the control of the Central Bank.  Moreover, for each $i \in N$ we suppose that immediately
after the $i$th intervention there is
a bounded, random length of time $T_i$, $0 \leq T_i \leq \bar T$, during which the exchange rate follows a new drift $\mu_2^i$
and volatility $\sigma_2^i$ (also Lipschitz), after which the drift and volatility revert to the original values. Here $\bar T$
is some positive uniform upper bound on the reaction times.

The new drift and volatility are also permitted to be random; their distributions
can be specified quite freely, as long as the sequences $\sigma_2^i$, $\mu_2^i$
and $T_i$, are $iid$ and independent of each other and of the rate process driver $W_t$.
For definiteness, we can specify the new drift and volatility rates as
\begin{equation}
\mu_2^i(x) = \mu_2^0(x) + \mu_{\delta}^ix \mbox{ and } \sigma_2(x)^i = \sigma_2^0(x) + \sigma_{\delta}^ix,
\end{equation}
where $\mu_2^0$ and $\sigma_2^0$ are fixed, known functions, and $\mu_{\delta}^i$ and $\sigma_{\delta}^i \geq 0$
are $iid$ with some known bounded probability distributions.
The case $\mu_{\delta}^i = 0 = \sigma_{\delta}^i$ for all $i$ corresponds to a market reaction process known
in advance;
the case $T^i=0$ for all $i$ corresponds to no market reaction, so we will henceforth assume $E[T_i] >0$.
 
 We denote by $ \tilde{X}_{y}^i(t) $ the diffusion process followed during this
 reaction period of duration $T_i$ after the $i$th intervention:
\begin{equation}
	d \tilde{X}_{y}^i(t) = \mu_2^i(\tilde{X}_{y}^i(t)) \  dt + \sigma_2^i(\tilde{X}_{y}^i(t)) \  dW_t \nonumber
\end{equation}
where $y$ is the value of the rate process immediately after intervention.

We impose the important restriction that new interventions are not allowed during
this reaction period.  Interventions are also restricted
to times $\tau \geq 0$, but the case $\tau = 0$ is allowed.  Our convention will be that
the controlled process is right continuous with left limits (cadlag), but to make sense
of the left limit at zero, we need the process defined in some neighborhood of zero,
so our time domain is $(-a, \infty)$, $a>0$, as mentioned above.

Given the uncontrolled process $X_y$, we now define an impulse control $\nu$ and it's corresponding cadlag controlled process $X_y^\nu$ as follows:
\begin{defn} \label{eqn:2reg_process}
An impulse control (strategy) $\nu= \left( \tau_1, \tau_2, ...; \xi_1,\xi_2,....\right)$ corresponding to the controlled
process $X_y^{\nu}$ is a sequence of intervention times
$\tau_i$ and control actions $\xi_i$, corresponding to a cadlag process $X_y^{\nu}(t)$ defined for $t \in (-a, \infty)$, such that
\begin{itemize}
\item $0 \leq \tau_i \leq \tau_{i+1}$ a.s. for all $i \in N$, and $\tau_i < \tau_{i+1}$ a.s. if $\tau_i < \infty$
\item $\tau_i$ is a stopping time with respect to the
filtration $\mathcal{F}^\nu_t = \sigma\{X_y^\nu(s-), s \leq t\}$, 
\item $\xi_i : \Omega \rightarrow \R$ is $\mathcal{F}_{\tau_i}$-measurable (Intuitively, $\tau_i$ indicates the time of the $i^{th}$ intervention of size $\xi_i$.)
\item $X^{\nu}_y(0-) = y$
\item $dX_y^\nu (t)   =   dX_y(t)$,   $t < \tau^{}_1$, and for all $j \in N$:
\item $X_y^\nu (\tau_j)  =   X_y^\nu (\tau_j-) - \xi_j$
\item $dX_y^\nu (t)   =   d\tilde{X}_{X_y^\nu (\tau_j)}^i(t)$,   $\tau_j \le t < \tau_j + T_j$
\item $dX_y^\nu (t)   =   d{X}_{X_y^\nu (\tau_j+T)}(t)$, $ \tau_j+ T_j \le t < \tau_{j+1}$
\end{itemize}
\end{defn}

We allow the sequence of intervention times to be of finite length $k$ by setting $\tau_i = \infty$ for $i>k$.
Also, it's possible that $\tau_{j+1} = \tau_j + T_j$, in which case the interval $\tau_j+ T_j \le t < \tau_{j+1}$ mentioned
in the last item of the definition is empty.  

We consider the performance function 
\begin{equation}
	J^\nu(x)= E \left[ \int_0^{\infty} e^{-rt} f(X^{\nu}_x(t)) dt + \sum_{i=1}^{\infty} e^{-r \tau_i}  K \left( X_x^\nu (\tau_j-), \xi_j    \right) \right] \nonumber
\end{equation}
where $K(x,\xi)$ is a given function that represents the cost of intervention depending on the state $x$ at intervention and the intervention size $\xi$. (In examples we often take $K(x,\xi)$ to be either
a constant $K$ or a constant plus proportional costs $K_1 + K_2|\xi|$.)
The constant $r$ represents the discount factor, assumed fixed here, and $f$ is a continuous running
cost function, for example measuring the deviation from a target value.

The Central Bank wants to use a policy that minimizes the performance function
over all possible admissible controls. Therefore, we define the Value Function as
\begin{equation}
	V(x)=\displaystyle \inf_{v \in \mathcal{V}} J^\nu(x) \nonumber
\end{equation}
where $\mathcal{V}$ is the set of admissible controls.  This value function depends on
the precise definition of admissible -- in application it is enough that
$\mathcal{V}$ includes all reasonable controls that might be considered in practice.
For our purposes we define $\mathcal{V}$ as follows.

\begin{defn} An impulse control
\begin{equation}
	\nu= \left( \tau_1, \tau_2, ...; \xi_1,\xi_2,....\right), \nonumber
\end{equation}
is {\bf admissible} ($\nu \in \mathcal{V}$) if 
\begin{equation}
X^{\nu}_x(\tau_i) > 0 \mbox{ for all $i \in N$, }
\label{eqn:adm0}
\end{equation}
\begin{equation}
\tau_{i+1} - \tau_i \ge T_i \mbox{ for all $i \in N$, } 
\label{eqn:2reg_adm2}
\end{equation}
\begin{equation}
E \left [ \int_0^\infty e^{-rt}f(X_x^\nu(t)) \ dt \right]  <  \infty,  \label{eqn:2reg_adm1}
\end{equation}
%\begin{equation}
%P \left( \displaystyle \lim_{i \rightarrow \infty} \tau_i \leq t \right)  =  0 \ \ \forall \ t \geq 0 ,  \label{eqn:2reg_adm2}
%\end{equation}
and
\begin{equation}
\displaystyle  E \left[ \int_0^{\infty} ( e^{-rt} X_x^\nu(t))^2 \, dt \right]  < \infty.  \label{eqn:2reg_adm3}
\end{equation}
\end{defn}

Condition (\ref{eqn:2reg_adm2}) means that the central bank will not intervene while the
market is still reacting to the previous intervention. This  implies that $\tau_n \to \infty$
as $n \to \infty$ almost surely: since the $T_i$ have positive mean, the law of large numbers
applies and $\tau_n \geq \sum_{i=1}^{n-1} T_i \to \infty$.
Conditions (\ref{eqn:2reg_adm1}) and
(\ref{eqn:2reg_adm3}) are mild boundedness conditions that will be easily satisfied by
any practical intervention policy and running cost function.

Additionally, let's define $\tilde{K}$ as the expected running cost immediately after an intervention takes place provided that the process restarts from $\bar{x}$ and that it remains under the second diffusion regime for a period of time $T$
\begin{equation} \label{eqn:Ktilde}
	\tilde{K}(\bar{x})=E \left[ \int_0^T e^{-rt} f(\tilde{X}_{\bar{x}}(t)) dt \right]. \nonumber
\end{equation}
(Since the sequences $(T_i)$, $(\mu_{\delta}^i)$, and $(\sigma_{\delta}^i)$ are $iid$ 
and independent of $\tilde X$, $\tilde{K}$ does not
depend on which intervention has occurred.  We write $T$ for a generic random variable
with the same distribution as $T_i$, and similarly for $\mu_{\delta}$, $\sigma_{\delta}$, and
$\tilde X$.)

%%%%%%%%%%%%%%%%%%%%%%%%%%%%%%%%%%%%%%%%%

\section{Quasi-variational Inequalities}

To solve the impulse control problem formulated above, 
We will use the quasi-variational inequalities (QVI) approach (Bensoussan and Lions \cite{bensoussan-lions84},
Korn \cite{korn99}), which involves
constructing the value function $V(x)$ as a solution to a system of inequalities described
below.  The value function $V(x)$ then determines the optimal control strategy, called
the QVI-control associated to $V$.  Once we have proved that solutions of the QVI yield the optimal intervention
strategy, this reduces the control problem to the much easier problem of solving the QVI.
We state the general results here and illustrate their use in the next section.

 First,
we require a new optimal intervention operator $\mathcal{M}$ adapted to our situation.
Let $\R^+ = \{ x \in \R: x > 0 \}$.

\begin{defn} For a function $\phi: \R^+ \to \R$, and $x \in \R^+ ,\xi \in \R$, define operators 
$\mathcal{M}$ and $M$ as follows: 
\begin{equation}
M(\phi, x, \xi) = K(x,\xi)+\tilde{K}(x-\xi) +  E\left[e^{-rT} \phi(\tilde{X}_{x-\xi}(T))\right]
\end{equation}
and
\begin{equation} \label{eqn:2reg_ModM}
\displaystyle \mathcal{M} \phi(x) = \inf_{\xi} \{ M(\phi, x, \xi): x-\xi > 0 \}.
\end{equation}
whenever these are well-defined.
\end{defn}

We also need the differential operator $\mathcal{L}$ given by
\begin{equation} \label{eqn:lambda1}
\mathcal{L}\phi = 
\frac{1}{2} \sigma_1^2(x)\frac{d^2}{d x^2}\phi + \mu_1(x) \frac{d}{d x}\phi - r\phi.
\end{equation}

This operator will be useful in two ways.  First, Ito's formula applied to any function
of the form $e^{-rt}\phi(X_x(t))$, where $X$ is the uncontrolled process (\ref{eqn:GBM}),
gives, for any two times $S < U$,

\begin{eqnarray} \label{eqn:Ito}
e^{-rt} \phi(X_x(t))|_{S}^{U} & = &  \int_S^{U} e^{-rt} \mathcal{L} \phi(X_x(t)) \ dt \\ & + & \int_S^{U} e^{-rt} \sigma_1 (X_x(t)) \phi'(X_x(t)) \ dW_t .
\end{eqnarray}

Second, when there are no interventions, 
$V$ takes the form
\begin{equation}
V(x) =   E\left[ \int_0^{\infty} e^{-rt} f(X_x(t)) dt \right]
\end{equation}
which, if $f$ is bounded and continuous (e.g. Oksendal \cite[Ch.~8]{oksendal03}), satisfies the equation 
\begin{equation}
\mathcal{L}V(x) + f(x) = 0.
\end{equation}

We will see that the solution of our impulse control problem splits the domain of $x$ in two regions, an intervention and a continuation region. In the intervention region, where it is optimal to intervene, it should be the case that $V(x)=\mathcal{M}V(x)$. On the other hand, the interval where intervention is not optimal because $\mathcal{M}V(x)>V(x)$, is referred to as the continuation region, and we have $\mathcal{L}V(x)+f(x)=0$ in that region. This suggests that the Value function should satisfy a set of inequalities, which are commonly known as the Quasi-Variational Inequalities.

\begin{defn}
We say that the function $\phi$ {\bf satisfies the quasi-variational inequalities (QVI)} if $\phi$
satisfies the following three conditions:
\begin{equation} \label{eqn:2reg_in1}
\mathcal{L}\phi(x) + f(x) \geq 0, 
\end{equation}
\begin{equation} \label{eqn:2reg_in2}
 \phi(x) \leq \mathcal{M} \phi(x),
\end{equation}
\begin{equation} \label{eqn:2ret_in3}
(\mathcal{L}\phi(x) + f(x)) (\phi(x) - \mathcal{M} \phi(x))=0.
\end{equation}
\end{defn}

From a solution of the quasi-variational inequalities, we construct the following control:
\begin{defn}
Let $\phi$ be a continuous solution of the QVI. Then the following impulse control is called a {\bf QVI-control associated to $\phi$} (if it exists):
\begin{eqnarray*}
      \tau_1 := &  \inf \{ t > 0: \phi(X_x(t-)) = \mathcal{M} \phi(X_x(t-)) \},  \mbox{ and for every } n > 1, \\
	\tau_n:= & \inf \{ t > \tau_{n-1} + T_{n-1} : \phi(X_x^\nu(t-))= \mathcal{M} \phi(X_x^\nu(t-)) \} \mbox{ and } \\
	\xi_n:= & \arg \min \bigg\{ K(X_x^\nu(\tau_n-),\xi) + \tilde{K}(X_x^\nu(\tau_n-) - \xi)  
	 + E  \left[ e^{-rT} \phi(\tilde{X}_{X_x^{\nu}(\tau_n-) - \xi}(T) | {\cal{F}}_{{{\tau}_n}-} \right] \\
	            &  : \xi \in \R, X_x^{\nu}(\tau_n-) - \xi > 0 \bigg\}. \nonumber
\end{eqnarray*}
\end{defn}
The $\arg \min$ above might not be unique, so we do not claim there is only one QVI-control -- though
in practice we do observe uniqueness.

The following main theorem permits us to verify that a solution of the QVI and the admissible control attached to it solve the impulse control problem.  Denote by
$\mathcal{L}_2$ the operator defined by
\begin{equation}\label{eqn:lambda2}
\mathcal{L}_2 \phi (x) = 
\frac{1}{2} \sigma_2^2(x)\frac{d^2}{d x^2} \phi(x) + \mu_2(x) \frac{d}{d x}\phi(x) - r\phi(x). 
\end{equation}

\begin{theorem} \label{thm:mythem}
Let $\phi \in C^1(\Rp)$ be a solution of the QVI and suppose there is a finite subset
$\mathcal{N} \subset \Rp$ such that $\phi \in C^2(\Rp - \mathcal{N})$. If $\phi$ satisfies the growth conditions
\begin{equation} \label{eqn:K1}
\displaystyle E \int_0^\infty (e^{-rt} \sigma_i(X_x^\nu(t)) \phi'(X_x^\nu(t)))^2 \ dt < \infty, 
\quad i = 1,2,
\end{equation}
\begin{equation} \label{eqn:K2}
\displaystyle \lim_{t \rightarrow \infty} E \left [ e^{-rt} \phi(X_x^\nu(t)) \right] =0,
\end{equation}
and
\begin{equation} \label{eqn:K3} 
\displaystyle  E \left [ \int_0^{\infty} e^{-rt} |\mathcal{L}_2 \phi(X_x^\nu(t))| \, dt \right] < \infty,
\end{equation}
for every process $X_x^\nu(t)$ corresponding to an admissible impulse control $\nu$, then for every $x \in \Rp$ 
\begin{equation}
	V(x) \geq \phi(x). \nonumber
\end{equation}
Moreover, if the QVI-control corresponding to $\phi$ is admissible then it is an optimal impulse control, and for every $x \in \Rp$
\begin{equation}
	V(x)= \phi(x).
\end{equation}
\end{theorem}

The theorem is proved in Section \ref{proof}.

%%%%%%%%%%%%%%%%%%%%%%%%%%%%%%%%%%%%%%%%%%%%%%%%%%%%
\section{Solving the QVI: An Illustration} \label{sec:example}

We now illustrate how theorem \ref{thm:mythem} is used in the context of finding the Central
Bank's optimal impulse control strategy.  Our results can be applied to quite general exchange
rate processes, but for simplicity we illustrate their use for the common case of geometric
brownian motion.  We can also handle quite general cost functions $K(x, \xi)$; to simplify our
examples we now restrict attention to the special case of fixed costs $K(x, \xi) = K$.

  As a warm-up, we first describe the known case
$T=0$ when there is no market reaction period. 
\subsection{The case without market reaction}

A Central Bank desires to keep the exchange rate close to a target $\rho$ using an impulse
control strategy.  The admissible controls are the same as above, except that since $T=0$
here
we need to replace
condition (\ref{eqn:2reg_adm2}) with the admissibility condition
\begin{equation}
P \left( \displaystyle \lim_{i \rightarrow \infty} \tau_i \leq t \right)  =  0 \ \ \forall \ t \geq 0 .  
\end{equation}

 Let $X_t$ represent domestic currency units per unit of foreign currency at time $t$ and suppose that the dynamics of $X_t$ are given by
$$X^{\nu}_x(t) = x + \int_0^t \mu X^{\nu}_x(s) \ ds + \int_0^t \sigma X^{\nu}_x(s) \ dW_s + \sum_{i=1}^\infty 1_{\{\tau_i < t \}} \xi_i ,$$
where $\nu = (\tau_i, \xi_i)_{i=1}^{\infty}$ is an admissible impulse control. This means
that $X^{\nu}_x(t)$ follows a geometric Brownian motion in the absence of interventions.

The Central Bank wants to find the optimal impulse control that minimizes the following functional that depends on the control $\nu$
$$J^{\nu}(x) = E \left [ \int_0^\infty e^{-rt} f(X^{\nu}_x(t)) \ dt + \sum_i e^{-r \tau_i} g(\xi_i) 1_{\{\tau_i<\infty\}}  \right], $$
where
$$f(x)= (x- \rho)^2,$$ and for ease of exposition in this illustration we take $g(\xi_i)= K$, for some positive constant $K$, meaning that there are only fixed intervention costs.
The Value Function for this example is
$$V(x)= \inf_{\nu} J^{\nu}(x)$$
where the infimum is taken over all admissible controls. 

Cadenillas and Zapatero \cite{cadenillas-zapatero99} show that the optimal control consists in forcing an intervention each time the exchange rate process hits the boundary of a band $[a,b]$, and the optimal intervention consists in jumping to a value $\alpha$, where $a< \alpha < b$. 
(When there are proportional intervention costs as well, they show that the strategy is to jump to the boundary
of a band properly contained in $[a,b]$; we observe similar results when $T>0$.)

They show that the Value Function is
\begin{equation}
V(x)= \nonumber
\begin{cases}
\phi(\alpha) + K  & if \ x<a \\
\phi(x) = Ax^{\gamma_1} + B x^{\gamma_2}+\left(\frac{1}{r-\sigma^2 -2 \mu}\right)x^2- \frac{2 \rho}{r-\mu}x + \frac{\rho^2}{r}  & if \ a \leq x \leq b \\
\phi(\alpha) + K & if \ x>b 
\end{cases}
\end{equation}
where $ \gamma_{1,2}= \frac{-\mu + 0.5 \sigma^2 \pm \sqrt{(\mu - 0.5 \sigma^2)^2+ 2 r \sigma^2   } }{\sigma^2 }$. The unknown parameters $A,B,a,\alpha, b$ are found using continuity, optimality, and smooth pasting conditions (as we will explain in the next subsection). Note that $V$ solves $\mathcal{L}V(x)+f(x)=0$ in the continuation region, $a<x<b$. In the intervention region, as there are no intervention costs, we have that $\mathcal{M}V(x)=V(\alpha)+K$.

%%%%%%%%%%%%%%%%%%%%%%%%%%%%%%%%%%%%%%%%%

\subsection{The case with market reaction}

We assume for this example that we have the same intervention and running costs as above,
but now after the $i$th intervention the volatility parameter changes to
$\sigma_2^i = \sigma + \sigma_{\delta}^i$ for a time $T_i>0$, representing a temporary new market
regime in reaction to the intervention. During this reaction time additional intervention is not
allowed.

Our method for finding the optimal impulse control strategy is to propose the form of
the optimal control $\nu$ (up to some unknown parameters), and use $\nu$ to construct
a solution $\phi$ to the QVI for which $\nu$ is the QVI-control. If, for the proper choice
of parameters, $\phi$ and $\nu$ satisfy
the smoothness and growth conditions of theorem \ref{thm:mythem}, and if $\nu$ is admissible,
we will be able to conclude by theorem \ref{thm:mythem} that $\phi$ is the value function for the problem and $\nu$ is our
desired optimal strategy.

The proposal is that we should intervene each time the exchange rate process leaves an interval $(a,b)$,
as soon as at least time $T$ has elapsed since the last intervention, and that the optimal intervention consists in shifting the exchange rate to $\alpha$, where  $a<\alpha<b$. 
The constants $a, b,$ and $\alpha$ are as  yet unknown.

If indeed it is optimal to intervene only outside the interval $(a,b)$, then we will expect $\phi$ to satisfy the differential equation $\mathcal{L} \phi(x) + f(x) = 0$ inside the interval. (The interval $(a,b)$ is called
the ``continuation region" because the exchange rate freely follows the original SDE in this interval,
after any reaction period.)

Therefore, solving $\mathcal{L} \phi(x) + f(x) = 0 $
we obtain, for $a < x < b$, 
\begin{equation}
 \phi(x) = Ax^{\gamma_1} + B x^{\gamma_2}+\left(\frac{1}{r-\sigma^2 -2 \mu}\right)x^2- \frac{2 \rho}{r-\mu}x + \frac{\rho^2}{r} \nonumber
\end{equation}
where $ \gamma_{1,2}= \frac{-\mu + 0.5 \sigma^2 \pm \sqrt{(\mu - 0.5 \sigma^2)+ 2 r \sigma^2   } }{\sigma^2 }$, and $A$ and $B$ are yet to be found.

We need to compute the running cost $\tilde{K}$ incurred during the reaction period $T$:
$$\tilde{K}(x)=E\left[ \int_0^T e^{-rt}(\tilde{X}_x(t)-\rho)^2 \ dt \right].$$
For fixed $T$ and $\sigma_2$, the above integral can be computed analytically because
$\tilde X_x(t)$ follows a Geometric Brownian Motion.  Since $T$ and $\sigma_2$ are
independent of $W_t$, these variables can be integrated separately:

\begin{eqnarray}
	\tilde{K}(x) & = & E \left[ \int_0^T e^{-rt}  (x^2e^{2(\mu t -\frac{1}{2} \sigma^2_2 t + \sigma_2 W_t)} -2 e^{-rt} (e^{(\mu t -\frac{1}{2} \sigma^2_2 t + \sigma_2 W_t)} \rho x + \rho^2 ) \ dt \right] \nonumber \\
	&=& E \left[ \int _{0}^{T}\!{e^{-rt}} \left( {x}^{2}{e^{2\,\mu\,t+{ \sigma^2_2}\,t
}}-2\,\rho\,x{e^{\mu\,t}}+{\rho}^{2} \right) {dt} \right] \nonumber 
\end{eqnarray}
where we have used $E[e^{\sigma_2 W_t}]=e^{\frac{1}{2}\sigma^2_2 t}$.

The boundary conditions at the intervention points $a,b$ can be found using continuity
and the condition $\phi = \mathcal{M}\phi$,
with the operator $\mathcal{M}$ defined by equation (\ref{eqn:2reg_ModM}), and
$\alpha$ denoting a minimizer in the definition of $\mathcal{M}$:
$$
\alpha = \arg \min \left(K + \tilde{K}(\alpha) + E [e^{-rT} \phi(\tilde{X}_{\alpha}(T)) ] \right).
$$

 We obtain
 \begin{equation} \label{eqn:sol1}
\phi(a) = \phi(b)  =   K + \tilde{K}(\alpha) + E \left[e^{-rT} \phi(\tilde{X}_{\alpha}(T)) \right]   
\end{equation}
where the expectation  $E \left[e^{-rT} \phi(\tilde{X}_{\alpha}(T)) \right]$ can be computed
by separating the independent variables as
$$
E\left[e^{-rT} \bigg\{ \int_a^b \phi(x) p(x;\alpha, T, \sigma_2) \ dx 
%  +  \phi(a) \int_{-\infty}^a p(x;\alpha,T,\sigma_2) \ dx + \phi(b) \int_b^{\infty} p(x;\alpha,T,\sigma_2) \ dx \bigg\}\right]
+ \phi(a) \int_{[a,b]^c} p(x;\alpha,T,\sigma_2) \ dx \bigg\}\right],
$$
where $[a,b]^c$ denotes the complement of $[a,b]$, and $p(x;\alpha,T,\sigma_2)$ is the probability density of the Log-Normal distribution of $\tilde{X}_{\alpha}(T)$:
$$\log(\tilde{X}_{\alpha}(T)) \sim N(\ln{\alpha} +\mu T - \frac{1}{2}\sigma_2^2T,  \sigma_2^2 T),$$
and the expectation $E$ is now over $T$ and $\sigma_2$.
%A similar equation applies for $\phi(b)$,
%\begin{equation} \label{eqn:sol2} 
%\phi(b)  =   K + \tilde{K}(\alpha) + E \left[ e^{-rT}\phi(\tilde{X}_{\alpha}(T)) \right] .
%\end{equation}

In addition we have the ``smooth pasting requirement" (the $C^1$ condition), needed for the application of theorem \ref{thm:mythem}
\begin{equation}
\phi'(a)=0, \label{eqn:sol3}
\end{equation}
\begin{equation}
\phi'(b)=0, \label{eqn:sol4}
\end{equation}
and the optimality of $\alpha$,
which implies solving 
\begin{equation}
\frac{d}{d \alpha} \left( \tilde{K}(\alpha) + E\left[e^{-rT} \int_{-\infty}^{+\infty} p(x;\alpha,T,\sigma_2) \phi(x) \ dx\right]  \right)=0, \label{eqn:sol5}
\end{equation}
where the expectation here is over the distribution of $T$ and $\sigma_2$.

The above equations can be solved to obtain the parameters $A, B, a,b$ and $\alpha$, and the Value function has the same structure as in the case without market reaction, but with these different parameters:
\begin{equation}
V(x)= \nonumber
\begin{cases}
K  + \tilde{K}(\alpha) + E [e^{-rT} \phi(\tilde{X}_{\alpha}(T)) ] & if \ x<a \\
\phi(x) = Ax^{\gamma_1} + B x^{\gamma_2}+\left(\frac{1}{r-\sigma^2 -2 \mu}\right)x^2- \frac{2 \rho}{r-\mu}x + \frac{\rho^2}{r}  & if \ a \leq x \leq b \\
K + \tilde{K}(\alpha) + E [ e^{-rT} \phi(\tilde{X}_{\alpha}(T)) ] & if \ x>b 
\end{cases}
\end{equation}

Note that by construction this function is continuous at $a$ and $b$, and we denote
the common value by
$$
\Theta := \phi(a) = \phi(b) = K  + \tilde{K}(\alpha) + E [e^{-rT} \phi(\tilde{X}_{\alpha}(T)) ].
$$

Finally, to complete the solution, we must verify the hypotheses of theorem \ref{thm:mythem} to conclude that this is indeed the Value Function, and that the proposed policy is optimal:

\begin{theorem}
Let $A,B,a,b,\alpha$ be a solution of the system of equations (\ref{eqn:sol1}) - (\ref{eqn:sol5})
with $a < \alpha < b$,
and let $\Theta$ be the constant defined above. Consider the $C^1$ function $V(x):(0,\infty) \rightarrow [0,\infty)$ defined by
\begin{equation}
V(x)= \phi(x) = Ax^{\gamma_1} + B x^{\gamma_2}+\left(\frac{1}{r-\sigma^2 -2 \mu}\right)x^2- \frac{2 \rho}{r-\mu}x + \frac{\rho^2}{r}
\end{equation}
for $a < x < b$, and $V(x) = \Theta$ otherwise.

If 
\begin{equation} \label{eqn:cond}
	a< \rho - (r\Theta)^{1/2} \mbox{ and } b> \rho + (r\Theta)^{1/2},
\end{equation}
and
\begin{equation} \label{eqn:valpha}
V(\alpha) < \Theta,
\end{equation}
then $V(x)$ is the Value Function; namely,
$$V(x)= \inf_{\nu} J^{\nu}(x)$$
and the optimal policy is the QVI-control corresponding to $V$, given by 
\begin{equation}\nonumber
\tau_1 = \inf \{ t > 0: X(t) \notin (a,b) \},
\end{equation}
\begin{equation}\nonumber
\tau_i=\inf \{ t>=\tau_{i-1}+T_{i-1}: X^\nu(t) \notin (a,b) \} \quad (i > 1)
\end{equation}
and
\begin{equation}\nonumber
X^\nu(\tau_i)=X^\nu(\tau_i-) - \xi_i = \alpha.
\end{equation}
\end{theorem}

\begin{proof}
We start by showing that $V$ satisfies the QVI:

\noindent{First QVI Inequality:}  From the definition of $V$,
\begin{equation*}
\mathcal{L}V(x)+f(x) = 
-r \Theta + (x-\rho)^2
\end{equation*}
when $x < a$ or $x > b$, and
\begin{equation*}
\mathcal{L}V(x)+f(x) = \mathcal{L} \phi(x) + (x-\rho)^2
\end{equation*}
when $a \le x \le b$.

By construction of $\phi$, $\mathcal{L} \phi(x) + (x-\rho)^2=0$ in the interval $[a,b]$; outside the interval we have that $\mathcal{L} \phi(x) + (x-\rho)^2>0$ because of conditions  (\ref{eqn:cond}).
\smallskip

\noindent{Second QVI Inequality:}
From the definitions of $\mathcal{M}$ and $\alpha$, we have
 $\mathcal{M}V(x)= K + \tilde{K}(\alpha)+ E[e^{-rT}V(\alpha)] = \Theta$ for all $x$. Therefore, $\mathcal{M}V(x)=V(x)$ for $x \notin [a,b]$. 
We need to show that $V(x) \leq \Theta$ for all $x \in (a,b)$, from which we can conclude
$\mathcal{M}V(x) \leq V(x)$.

To see this, notice that on the interval $(a,b)$, the third derivative $V'''$ is of the form $c_1x^{d_1} + c_2x^{d_2}$,
which can have at most one zero on $\R^+$.  Therefore the second derivative $V''$ can change sign at most
twice in $(a,b)$.  Because $V(a)=V(b)= \Theta$ and $V'(a)=V'(b)=0$, this means $V(x) - \Theta$ cannot take
both signs on $[a,b]$ -- this would imply at least three sign changes for the second derivative -- so we must either
have $V(x) \leq \Theta$ or $V(x) \geq \Theta$ for all $x$.  Because of (\ref{eqn:valpha}), it must be the former.

\noindent{Third QVI Inequality:}
Holds as a result of the above arguments.

To verify the remaining conditions of theorem \ref{thm:mythem},  first note that $V'(x)$ is bounded, so condition (\ref{eqn:K1}) holds via the admissibility condition (\ref{eqn:2reg_adm3}). Condition (\ref{eqn:K2}) is immediately satisfied because $V$ is bounded. Condition (\ref{eqn:K3}) is satisfied because $V''$ is bounded,
even though discontinuous at $a$ and $b$.

Finally, the QVI control given above is admissible:  condition (\ref{eqn:adm0}) holds since $X$ is lognormal
and $\alpha > 0$; condition (\ref{eqn:2reg_adm2}) holds by construction; and conditions (\ref{eqn:2reg_adm1})
and (\ref{eqn:2reg_adm3}) hold because $X$ and $\tilde X$ are lognormal and $X^{\nu}$ is bounded inside $[a,b]$
except for possible excursions for a duration at most $\bar T$.

\end{proof}

\subsection{Numerical Example:}
Consider $\rho=1.4$, $r=0.06$, $\mu=0.1$, $\sigma=0.3$, a fixed $\sigma_2=0.4$, $K=0.5$, and a fixed recovery time $T=1$. 

A searching algorithm was implemented to obtain the parameters $A,B,a,b$ and $\alpha$ that solve the system of equations (\ref{eqn:sol1}) - (\ref{eqn:sol5}). The key element that allows solving the expectation $E \left[e^{-rT} \phi(\tilde{X}_{\alpha}(T)) \right]$ in equations (\ref{eqn:sol1}) and (\ref{eqn:sol5}) is realizing that $\phi(x)$ is constant outside $(a,b)$. For fixed $T$, the expectation can then be expressed as:
$$
 e^{-rT} \bigg\{ \int_a^b \phi(x) p(x;\alpha,T,\sigma_2) \ dx
  +  \phi(a) \int_{-\infty}^a p(x;\alpha,T,\sigma_2) \ dx 
  $$
  $$ + \phi(b) \int_b^{\infty} p(x;\alpha,T,\sigma_2) \ dx \bigg\}.
$$
These integrals can be computed for a given trial set of parameters $A,B,a,b$ and $\alpha$. 
(The evaluation of the expectation for the case of fixed and proportional transactions costs is also possible with some modifications.)

\begin{figure}[ht]
\centering
\includegraphics[width=0.90\textwidth]{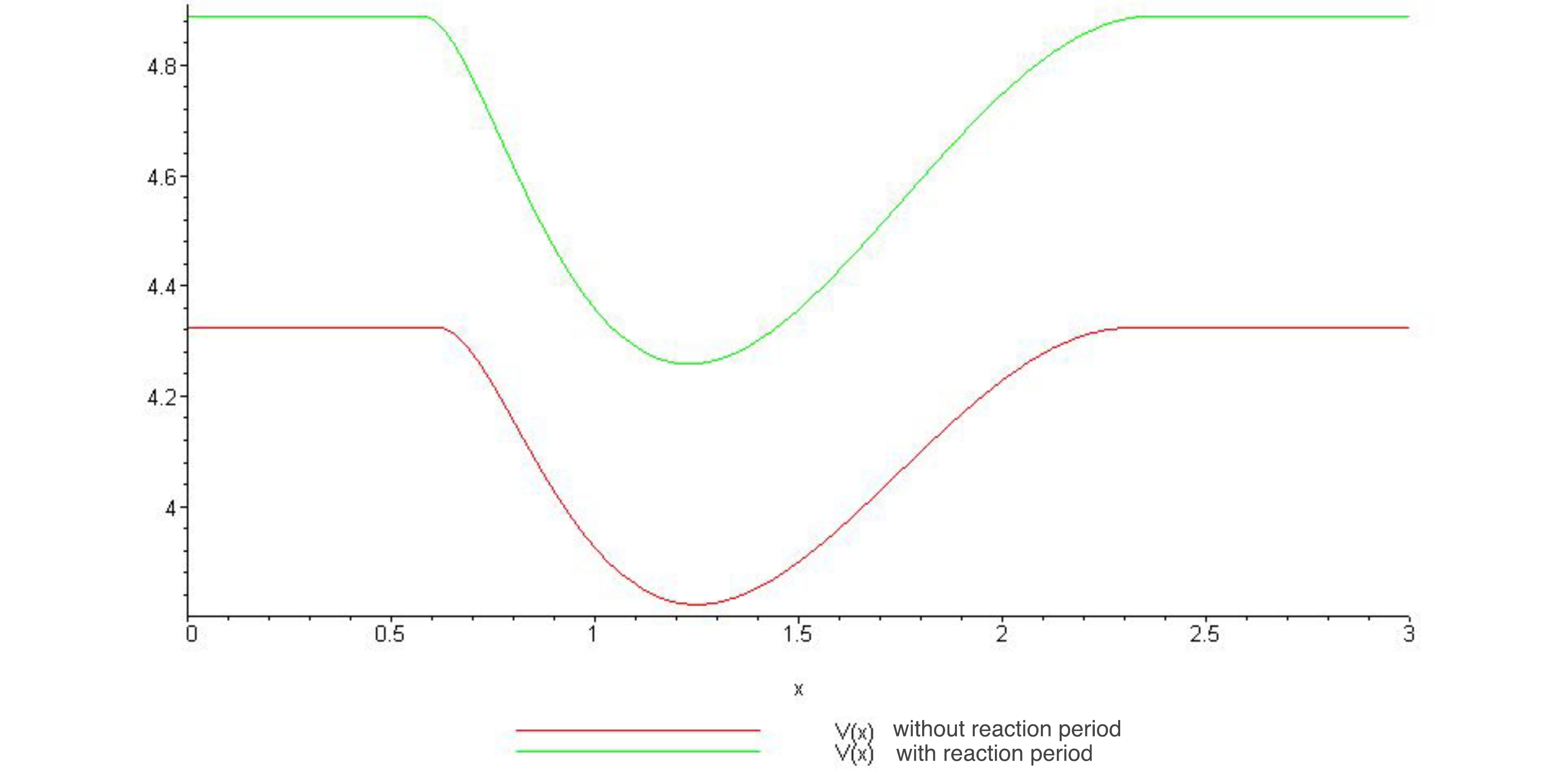}
\caption{Value function with one and two volatility regimes. The optimal strategy is more expensive if there are two volatility regimes and a positive reaction period.}
\label{fig:2reg_fig5}
\end{figure}

Table \ref{tab:2reg_tab2} and figure \ref{fig:2reg_fig5} compare the policy results with and without the reaction period. The coefficients $A$ and $B$ are found to be $-1.330$ and $-93.064$ when $T=0$ and $-1.193$ and $-92.759$ when $T=1$. (Conditions (\ref{eqn:cond}) are easily verified.)

The existence of a new regime after interventions requires a modification in the policy observed by the Central Bank: the intervention points are different, notice that the band is widened; in addition, the amount of intervention is also different, as the new restarting value $\alpha$ indicates.  The presence of a market reaction leads to the need
for greater patience by the Central Bank, less frequent but larger interventions, and greater optimal costs.

We also observe that the effect of a reaction period in the model, even with a deterministic reaction time and amount
as in this numerical example, cannot be captured simply by adjusting the cost function instead. In this
example, by changing $K$ from 0.5 to 0.63, and keeping reaction time $T=0$, we can match the intervention points
$a$ and $b$ obtained when $T=1$, but the optimal restarting value is different: $\alpha = 1.230$ instead of $1.212$;
see table \ref{tab:2reg_tab2}.  The introduction of a market reaction in the model leads to a different policy than
would arise by simply adjusting the intervention cost upward to compensate for the increased uncertainty.

\subsection{Comparative Statics Analysis}
	
One of the main advantages of obtaining analytical solutions is that comparative analysis can be performed. If a Central Bank knows what type of reaction to expect from the market after performing interventions -- namely, an increase (decrease) in volatility, or an increase (decrease) in the trend on the dynamics of its currency -- then the optimal policy can be found.

In table \ref{tab:2reg_tab3} we present the optimal policy in four different scenarios. Both the bands and the intervention sizes depend on the nature of the reaction of the market during the reaction time after interventions. It is reasonable to suppose volatility increases after interventions and, as we showed above, this implies a wider band than in the case without market reaction. For comparison purposes we also computed the case when the volatility decreases during the reaction period; in this case the band shrinks. It could also be the case that the trend of the currency is modified as a temporary outcome of the intervention; perhaps reflecting the sentiment of the market with respect to the confidence on the Central Bank's actions. It is interesting to observe that the band widens if the drift trend increases or decreases and, as expected, the optimal restarting point is closer to the long term target $\rho=1.4$ when the drift is lowered during the reaction period. 

In table \ref{tab:2reg_tab4} we show the effect of varying $T$ for the case when the market reaction is a temporary volatility increase to $\sigma_2=0.4$. Observe that as $T$ increases the band widens. We conclude the analysis with the optimal policy when the reaction time $T$ is uniformly distributed between $0$ and $1$.

%%%%%%%%%%%%%%%%%%%%%%%%%%%%%%%%%%%%%%	%%%%%%%%%%%%%
	\section{Proof of Theorem \ref{thm:mythem}}
	\label{proof}
	
Before we prove the theorem, we need the following lemma:

\begin{lemma} \label{lem:milema}
Let $\nu$ be an admissible control, and let $\sigma $ be an intervention time for $\nu$.
If the function $\phi$ satisfies $\phi \le \mathcal{M}\phi$ and the growth condition (\ref{eqn:K1}), then we have the inequality
\begin{equation*}
e^{-r \sigma}\left( \phi(X_x^{\nu}(\sigma-)) -  \phi(X_x^{\nu}(\sigma)) \right)  
\end{equation*}
\begin{equation*}
 \leq e^{-r \sigma} K +   \displaystyle E \left[ \int_\sigma^{\sigma+T} e^{-rt} (\mathcal{L}_2 \phi(X_x^{\nu}(t))+ f(X_x^{\nu}(t))\ dt  | \mathcal{F}_{\sigma} \right], \nonumber
\end{equation*}
with equality if $\phi = \mathcal{M}\phi$, and where $\mathcal{L}_2$ is the operator defined by
\begin{equation}
\frac{1}{2} \sigma_2^2(x)\frac{d^2}{d x^2} + \mu_2(x) \frac{d}{d x} - r. 
\end{equation}
\end{lemma}

\begin{proof}
Application of Ito's formula (\ref{eqn:Ito}) to $e^{-rt}\phi(X_x^{\nu}(t))$ gives
\begin{eqnarray*}
e^{-r(\sigma+T)} \phi(X_x^{\nu}(\sigma+T)) & = & e^{-r \sigma}\phi(X_x^{\nu}(\sigma)) + \int_\sigma^{\sigma+T} e^{-rt} \mathcal{L}_2 \phi(X_x^{\nu}(t)) \ dt \\ 
& + & \int_\sigma^{\sigma+T} e^{-rt} \sigma_2 (X_x^{\nu}(t)) \phi'(X_x^{\nu}(t)) \ dW_t,  
\end{eqnarray*}
because between times $\sigma$ and $\sigma+T$ the controlled process follows the second diffusion. Taking conditional expectations we obtain
\begin{equation} \label{eqn:temp_lem1}
E \left[ e^{-r(\sigma+T)} \phi(X_x^{\nu}(\sigma+ T)) \ | \ \mathcal{F}_{\sigma} \right]  =   e^{-r \sigma}\phi(X_x^{\nu}(\sigma)) 
\end{equation}
\begin{equation*}
+ E \left[ \int_\sigma^{\sigma+T} e^{-rt} \mathcal{L}_2 \phi(X_x^{\nu}(t)) \ dt    
 +    \int_\sigma^{\sigma+T} e^{-rt} \sigma_2 (X_x^{\nu}(t)) \phi'(X_x^{\nu}(t)) \ dW_t \ | \ \mathcal{F}_{\sigma} \right]
 \end{equation*}
 \begin{equation*}
  =	 e^{-r \sigma}\phi(X_x^\nu(\sigma))  +  
  E \left[ \int_\sigma^{\sigma+T} e^{-rt} \mathcal{L}_2 \phi(X_x^\nu(t)) \ dt \ | \ \mathcal{F}_{\sigma} \right] 	
\end{equation*}
where the stochastic integral vanishes by condition
(\ref{eqn:K1}) and that fact that the Ito integral is a martingale. 

Now, the inequality
$\phi \le \mathcal{M}\phi$ says that for any positive $z$ and $y$,
\begin{eqnarray}
\phi(z) -  E \left[ e^{-rT}\phi(\tilde X_y^\nu(T)) \right] & \leq & K + \tilde{K}(y) \nonumber \\
& = & K + E\int_0^T e^{-rt} f(\tilde X_y^\nu(t)) \ dt , \nonumber		
\end{eqnarray}

Therefore, using $z=X_x^{\nu}(\sigma-)$, $y=X_x^{\nu}(\sigma)$,  this equation can be written as
\begin{equation*}
\phi(X_x^\nu(\sigma-)) -  E \left[ e^{-rT}\phi(X_x^\nu(\sigma+T)) \ | \ \mathcal{F}_{\sigma} \right]
\end{equation*}
\begin{equation*}
 \leq  K + E \left [ \int_{\sigma}^{\sigma+T} e^{-r(t-\sigma)} f(X_x^\nu(t)) \ dt \ | \ \mathcal{F}_{\sigma} \right]
\end{equation*}
Multiplying by $e^{-r\sigma}$ we obtain
\begin{equation*}
e^{-r \sigma}  \phi(X_x^\nu(\sigma-)) -  E \left[e^{-r(\sigma+T)} \phi(X_x^\nu(\sigma+ T))  |  \mathcal{F}_{\sigma} \right]  
\end{equation*}
\begin{equation*}
 \leq  e^{-r \sigma}K +  E \left [ \int_{\sigma}^{\sigma+T} e^{-rt} f(X_x^\nu(t))  dt  |  \mathcal{F}_{\sigma} \right], 
\end{equation*}
therefore using (\ref{eqn:temp_lem1}) we have
\begin{eqnarray} 
e^{-r \sigma}  \phi(X_x^\nu(\sigma-)) - \displaystyle e^{-r \sigma}\phi(X_x^\nu(\sigma))  & \leq &  \displaystyle E \left[ \int_\sigma^{\sigma+T} e^{-rt} \mathcal{L}_2 \phi(X_x^\nu(t)) \ dt \ | \ \mathcal{F}_{\sigma} \right] \nonumber \\
& + & e^{-r \sigma}K +  E \left [ \int_{\sigma}^{\sigma+T} e^{-rt} f(X_x^\nu(t)) \ dt \ | \ \mathcal{F}_{\sigma} \right], \nonumber
\end{eqnarray}
which is the desired result.
\end{proof}
%%%%%%%%%%%%%%%%%%%%%%%%%%%%%%%%%%%%%%%%%%%%%%%%%%%%%%%%%%%%%%%%%%%%%%%%%%%%%%%%%%%%%%%%

%%%%%%%%%%%%%%%%%%%%%%%%%%%%%%%%%%%%%%%%%%%%%%%%%%%%%%%%%%%%%%%%%%%%%%%%%%%%%%%%%%%%%%%%%%%%%
\vskip1cm
\noindent We now prove theorem \ref{thm:mythem}.
\medskip 

\noindent
\begin{proof}
Consider any admissible control $v= \{(\tau_n, \Delta X_n) \}_{n \in N}$. 

Define the stopping time $\tau^*(t) = \max \{ \tau_i : \tau_i \leq t \}$; note $\tau^*(t) \to \infty$ as
$t \to \infty$ almost surely since $\tau_i \to \infty$ a.s.

We need to estimate the quantity
\begin{equation} \label{eqn:CZ24_b}
	 e^{-r \tau^*(t)} \phi(X_x^{\nu}(\tau^*(t))) - \phi(x)  = A + B
\end{equation}
where $A$ and $B$ are the {\it finite} sums given by
\begin{eqnarray*}
	 A =&  1_{\{\tau_1 \leq t\}} (e^{-r\tau_1} \phi(X_x^\nu(\tau_1-)) - \phi(x)) + \\
	 &\sum_{i=2}^{\infty} 1_{\{ \tau_i \leq t \} } \left( e^{-r \tau_i}\phi(X_x^{\nu}(\tau_i-)) - 
	 e^{-r \tau_{i-1}}\phi(X_x^{\nu}(\tau_{i-1})) \right)
\end{eqnarray*}
and
\begin{equation*}
B =  \sum_{i=1}^{\infty} 1_{\{ \tau_i \leq t \} } e^{-r \tau_i} \left( \phi(X_x^{\nu}(\tau_i) - \phi(X_x^{\nu}(\tau_i-))   \right). 
\end{equation*}

For the terms in the summation in A, in the event $\{\tau_i \leq t\}$,
when $i = 2,3,\dots, n$, an application of Ito's formula gives
\begin{eqnarray}
e^{-r \tau_i} \phi(X_x^{\nu}(\tau_i-)) - e^{-r \tau_{i-1}} \phi(X_x^{\nu}(\tau_{i-1})) = & \displaystyle \int_{\tau_{i-1}}^{\tau_{i-1}+T} e^{-r s} \mathcal{L}_2\phi(X_x^{\nu}(s)) \ ds  \nonumber \\
+ & \displaystyle  \int_{\tau_{i-1}+T}^{\tau_i} e^{-r s} \mathcal{L}\phi(X_x^{\nu}(s)) \ ds \nonumber \\
+ & \displaystyle \int_{\tau_{i-1}}^{\tau_{i-1}+T} e^{-r s} \phi'(X_x^{\nu}(s)) \sigma_2(X_x^{\nu}(s)) \ dW_s \nonumber\\
 + & \displaystyle \int_{\tau_{i-1}+T}^{\tau_i} e^{-r s} \phi'(X_x^{\nu}(s)) \sigma_1(X_x^{\nu}(s)) \ dW_s. \nonumber
\end{eqnarray}
Here we are dropping the index $i$ from $T_i$ and $\sigma_2^i$ because they are $iid$ and independent of
$W_t$, and therefore the computations here are the same as if they were fixed.

Using QVI inequality (\ref{eqn:2reg_in1}), $\mathcal{L}\phi + f \geq 0$, this expression becomes
\begin{eqnarray}
e^{-r \tau_i} \phi(X_x^{\nu}(\tau_i-)) - e^{-r \tau_{i-1}} \phi(X_x^{\nu}(\tau_{i-1})) \geq 
& \displaystyle \int_{\tau_{i-1}}^{\tau_{i-1}+T} e^{-r s} \mathcal{L}_2\phi(X_x^{\nu}(s)) \ ds  \nonumber \\
+ &  \displaystyle \int_{\tau_{i-1}+T}^{\tau_i} e^{-r s} (-f(X_x^{\nu}(s)) \ ds \nonumber \\
+ & \displaystyle \int_{\tau_{i-1}}^{\tau_{i-1}+T} e^{-r s} \phi'(X_x^{\nu}(s)) \sigma_2(X_x^{\nu}(s)) \ dW_s \nonumber\\
+ &  \displaystyle \int_{\tau_{i-1}+T}^{\tau_i} e^{-r s} \phi'(X_x^{\nu}(s)) \sigma_1(X_x^{\nu}(s)) \ dW_s,\nonumber
\end{eqnarray}
with equality when $\mathcal{L}\phi + f = 0$.

For the term in A preceding the summation, we have, similarly,
\begin{eqnarray}
& e^{-r \tau_1} \phi(X_x^{\nu}(\tau_1-)) -   \phi(x)   \nonumber \\
&  = \displaystyle \int_{0}^{\tau_1}  e^{-r s}  \mathcal{L}\phi(X_x^{\nu}(s)) \ ds \nonumber 
 + \int_{0}^{\tau_{1}} e^{-r s}  \phi'(X_x^{\nu}(s)) \sigma_1(X_x^{\nu}(s)) \ dW_s, \nonumber \\
 \geq &   \displaystyle \int_{0}^{\tau_1}  e^{-r s}  (-f(X_x^{\nu}(s))) \ ds \nonumber 
 + \int_{0}^{\tau_{1}} e^{-r s}  \phi'(X_x^{\nu}(s)) \sigma_1(X_x^{\nu}(s)) \ dW_s. \nonumber
\end{eqnarray}

For the term in the second summation B of equation (\ref{eqn:CZ24_b}), we  use lemma \ref{lem:milema} so that in the event $ \{ \tau_i \leq t \}$ we have
\begin{equation*}
\displaystyle  e^{-r \tau_i} \left( \phi(X_x^{\nu}(\tau_i)- \phi(X_x^{\nu}(\tau_i-)) \right) 
\end{equation*}
\begin{equation*}
\geq - e^{-r \tau_i} K - E \left[ \int_{\tau_i}^{\tau_i+T} e^{-rs} \left( \mathcal{L}_2 \phi(X_x^{\nu}(s)) + f(X_x^{\nu}(s) \right) \ ds | \mathcal{F}_{\tau_i}\right],
\end{equation*}
with equality when $\phi = \mathcal{M}\phi$.

Therefore, reversing the sign and writing the $i=1$ and $i>1$ terms separately,
equation (\ref{eqn:CZ24_b}) becomes 
\begin{equation} \label{eqn:withis}
\phi(x) -  e^{-r \tau^*(t)} \phi(X_x^{\nu}(\tau^*(t))) \leq I_1 + I_2  
\end{equation}
where $I_1$ and $I_2$ correspond the terms $i=1$ and $i=2,\dots,n$; namely,
\begin{eqnarray*}
 I_1 = &  \displaystyle 1_{\{\tau_1 \leq t\}} \bigg( e^{-r \tau_1}K + E \left[ \int_{\tau_1}^{\tau_1+T} e^{-rs} (\mathcal{L}_2 \phi(X_x^{\nu}(s))+f(X_x^{\nu}(s))) \ ds | \mathcal{F}_{\tau_1}\right]   \nonumber \\
 + & \displaystyle \int_{0}^{\tau_{1}} e^{-rs} f((X_x^{\nu}(s))) \ ds  \nonumber 
 -  \displaystyle \int_{0}^{\tau_{1}} e^{-rs} \phi'(X_x^{\nu}(s)) \sigma_1(X_x^{\nu}(s)) \ dW_s \bigg) , 
\end{eqnarray*}
\begin{eqnarray*}
I_2 =& \displaystyle  \sum_{i=2}^{\infty} 1_{\{\tau_i \leq t  \}} \bigg\{   \left[ e^{-r \tau_i}K + E \left[ \int_{\tau_i}^{\tau_i+T} e^{-rs} (\mathcal{L}_2 \phi(X_x^{\nu}(s)) + f(X_x^{\nu}(s))) \ ds | \mathcal{F}_{\tau_i}\right] \right]  \nonumber \\
- &  \displaystyle \int_{\tau_{i-1}}^{\tau_{i-1}+T} e^{-rs} \mathcal{L}_2 \phi(X_x^{\nu}(s)) \ ds + \int_{\tau_{i-1}+T}^{\tau_i} e^{-rs} f(X_x^{\nu}(s)) \ ds \nonumber \\
- & \displaystyle \int_{\tau_{i-1}}^{\tau_{i-1}+T} e^{-rs} \phi'(X_x^{\nu}(s)) \sigma_2(X_x^{\nu}(s)) \ dW_s 
		-  \int_{\tau_{i-1}+T}^{\tau_i} e^{-rs} \phi'(X_x^{\nu}(s)) \sigma_1(X_x^{\nu}(s)) \ dW_s \bigg\} .  \nonumber
\end{eqnarray*}

We now take expectations on both sides of equation (\ref{eqn:withis})
\begin{equation} \label{eqn:withis2}
\phi(x) - E \left[ e^{-r \tau^*(t)} \phi(X_x^{\nu}(\tau^*(t))) \right] \leq E[I_1] + E[I_2], 
\end{equation}
and realize that the expectations of the stochastic integrals vanish because of condition (\ref{eqn:K1}).
 If we collect all terms of the right hand side where the integrand is $f(X_x^{\nu}(t))$, we have
\begin{eqnarray}
& \displaystyle  
E \left[ 1_{\{\tau_1 \leq t\}} \left( E\left[ \int_{\tau_1}^{\tau_1+T} e^{-rs} f(X_x^{\nu}(s))\ ds | \mathcal{F}_{\tau_1}\right]  +
   \int_{0}^{\tau_{1}} e^{-rs} f((X_x^{\nu}(s))) \ ds \right) \right]   & \nonumber \\
 + & \displaystyle  E \left[ \sum_{i=2}^{\infty}  1_{\{\tau_i \leq t  \}} \left(E \left [ \int_{\tau_i}^{\tau_i+T} e^{-rs} ( f(X_x^{\nu}(s))) \ ds | \mathcal{F}_{\tau_i}\right]  
  + \int_{\tau_{i-1}+T}^{\tau_i} e^{-rs} f(X_x^{\nu}(s)) \ ds  \right) \right] & \nonumber \\
 = & \displaystyle E \left[  \int_0^{\tau^*(t)+T} e^{-rs}f(X_x^{\nu}(s) \right], & \nonumber
\end{eqnarray}
where we have used 
 $$ 
 \displaystyle E \left[E[\int_{\tau_i}^{\tau_i+T} e^{-rs} ( f(X_x^{\nu}(s))) \ ds] \ | \ \mathcal{F}_{\tau_i}\right]=E \left[\int_{\tau_i}^{\tau_i+T} e^{-rs} ( f(X_x^{\nu}(s))) \ ds \right].
 $$
 
Collecting all terms where the integrand is $\mathcal{L}_2 \phi(X_x^{\nu}(t))$ in equation (\ref{eqn:withis2}) we obtain
\begin{eqnarray}
& \displaystyle 1_{\{\tau_1 \leq t\}}E \left(E \left[ \int_{\tau_1}^{\tau_1+T} e^{-rs} (\mathcal{L}_2 \phi(X_x^{\nu}(s))) \ ds | \mathcal{F}_{\tau_1}\right] \right)  \nonumber \\
 + & \displaystyle  \sum_{i=2}^{\infty} 1_{\{\tau_i \leq t  \}} \bigg\{  E \left( E \left[ \int_{\tau_i}^{\tau_i+T} e^{-rs} (\mathcal{L}_2 \phi(X_x^{\nu}(s)) ) \ ds | \mathcal{F}_{\tau_i}\right] \right)  \nonumber \\
-  & \displaystyle E \left[ \int_{\tau_{i-1}}^{\tau_{i-1}+T} e^{-rs} \mathcal{L}_2 \phi(X_x^{\nu}(s)) \ ds \right] \bigg\} \nonumber \\
= & \displaystyle E \left[\int_{\tau^*(t)}^{\tau^*(t)+T} e^{-rs}\mathcal{L}_2 \phi(X_x^{\nu}(s)) \ ds \right]. \nonumber
\end{eqnarray}

Hence, equation \ref{eqn:withis2} becomes
\begin{equation} \label{eqn:withis3}
\phi(x) - E \left[ e^{-r \tau^*(t)} \phi(X_x^{\nu}(\tau^*(t))) \right]
\end{equation}
\begin{equation*}
 \leq \displaystyle E [  \int_0^{\tau^*(t)} e^{-rs}f(X_x^{\nu}(s)) \ ds +  \int_{\tau^*(t)}^{\tau^*(t)+T} e^{-rs} (f(X_x^{\nu}(s)) +\mathcal{L}_2 \phi(X_x^{\nu}(s))) \ ds 
\end{equation*}
\begin{equation*}
+ \displaystyle  \sum_{i=1}^{\infty}  1_{\{\tau_i \leq t  \}} e^{-r \tau_i}K ].
\end{equation*}

%We now take the limit as $n \rightarrow \infty$. For the left hand side we obtain
%\begin{equation}
%	\displaystyle \lim_{n \rightarrow \infty} \left[ \phi(x) - E[ e^{-r \tau_n} \phi(X_x^{\nu}(\tau_n))  ]  \right] = \phi(x) - E[e^{-rt} \phi(X_x^{\nu}(t))], \nonumber
%\end{equation}
%and (\ref{eqn:withis3}) becomes
%\begin{equation*}
%\phi(x) - E [ e^{-rt} \phi( X_x^{\nu}(t))  ] 
%\end{equation*}
%\begin{equation*}
% \leq \displaystyle E \left[  \int_0^t e^{-rs}f(X_x^{\nu}(s)) \ ds +  \int_t^{t+T} e^{-rs}( f(X_x^{\nu}(s))+\mathcal{L}_2 \phi(X_x^{\nu}(s))) \ ds \right]  
% \end{equation*}
% \begin{equation*}
%+ \displaystyle E \left[ \sum_{i=1}^\infty  1_{\{\tau_i \leq t  \}} e^{-r \tau_i}K \right].
%\end{equation*}

Now, let $t$ go to infinity, notice that condition (\ref{eqn:K2}) tells us that the left-hand side becomes
$$
 \displaystyle \lim_{t \rightarrow \infty}  \phi(x) - E[ e^{-r\tau^*(t)} \phi(X_x^{\nu}(\tau^*(t))) ] = \phi(x),
 $$
while condition (\ref{eqn:K3}), the admissibility condition (\ref{eqn:2reg_adm1}), and
the dominated convergence theorem
 imply that  
 $$
 \displaystyle E \left[ \int_{\tau^*(t)}^{\tau^*(t)+T} e^{-rs} (f(X_x^{\nu}(s)) +\mathcal{L}_2 \phi(X_x^{\nu}(s))) \ ds \right] \rightarrow 0 \ \ as \ \ t \rightarrow \infty .
 $$ 

 Therefore,
$$ 
\phi(x) \leq E \left[ \sum_{i=1}^{\infty} 1_{\{\tau_i < \infty  \}} e^{-r \tau_i} K + \int_0^\infty e^{-rs} f(X_x^{\nu}(s)) \ ds  \right].
$$
Hence we have shown that
$$ \phi(x) \leq J^v(x).$$
As this is true for any control $v$, we have
$$\phi(x) \leq V(x).$$
Now, the above inequalities become equalities for the QVI-control associated to $\phi$.
\end{proof}	

	\section{Conclusions}
	We have addressed the problem of Central Bank intervention in the exchange rate market incorporating the effect of a temporary market reaction, of random duration,
affecting the dynamics of the exchange rate process. The reaction time $T$ can have any bounded
non-negative distribution provided it is stationary and independent of the rate process.
Using the Quasi-Variational Inequalities approach to impulse control problems, we presented a verification theorem that allows us to find the optimal policy and the Value Function for the problem. The main technical innovation is the use of a new optimal intervention
	operator $\mathcal{M}$ adapted to this setting.
	
We obtained an explicit solution of the problem for geometric brownian motion, and showed how the optimal policies are influenced by the presence of the reaction period after interventions. If volatility is assumed to
jump up temporarily after interventions, the result is that the target band widens and the optimal
costs increase.  The band narrows if interventions are assumed to cause a decrease in volatility;
a change of drift in either direction causes the band to widen slightly.  Thus, in most cases, Banks
should optimally intervene a little less often than they would if interventions were completely
invisible to the FX market.
	
It would be interesting to extend these results in various ways. Can the assumption that no
interventions are allowed during the reaction period be relaxed or removed?  What if the reaction
time is not independent of the process?  What if the drift and volatility during the reaction time are random
rather than fixed?  These are topics for further work.

\bibliographystyle{siam}
\bibliography{marketIntervention}  

\newpage

\newpage

\begin{table*}[htbp]
	\centering
		\caption{Optimal Policy with and without a recovery period. $K$ is the fixed intervention cost in this illustration,
		$a$ and $b$ are the lower and upper intervention levels,
		and $\alpha$ is the optimal restarting value.  }
\begin{tabular}{|l|c|c|c|l|}
\hline
Reaction Period & $K$ & $a$  & $b$ & $\alpha$ \\
\hline
None ($T=0$) & 0.5 & 0.622 & 2.307 & 1.249\\
$T=1$ &  0.5 &0.581 & 2.365 & 1.212\\
None ($T=0$) & 0.63 & 0.581 & 2.365 & 1.230\\
\hline
\end{tabular}
	\label{tab:2reg_tab2}
\end{table*}

\newpage

	\begin{table*}[htbp]
	\centering
		\caption{Optimal policy for different market reactions.}
\begin{tabular}{|l|c|c|l|}
\hline
Type of reaction & $a$  & $b$ & $\alpha$ \\
\hline
No reaction ($\sigma_1= \sigma_2=0.30$, $\mu_2=\mu_1=0.10$) & 0.622 & 2.307 & 1.249\\
Volatility increases ($\sigma_2=0.40$, $\mu_2=0.10$) & 0.581 & 2.365 & 1.212\\
Volatility decreases ($\sigma_2=0.10$, $\mu_2=0.10$) & 0.678 & 2.230 & 1.235\\
Drift increases ($\sigma_2=0.30$, $\mu_2=0.15$) & 0.618 & 2.314 & 1.186\\
Drift decreases ($\sigma_2=0.30$, $\mu_2=0.05$) & 0.621 & 2.309 & 1.275\\
\hline
\end{tabular}
	\label{tab:2reg_tab3}
\end{table*}

\newpage

\begin{table*}[htbp]
	\centering
		\caption{Optimal Policy for different reaction time periods when volatility increases during the reaction time.}
\begin{tabular}{|l|c|c|l|}
\hline
Reaction Period & $a$  & $b$ & $\alpha$ \\
\hline
None ($T=0$) & 0.622 & 2.307 & 1.249\\
$T=1$ & 0.581 & 2.365 & 1.212\\
$T=2$ & 0.516 & 2.461 & 1.072\\
$T \sim U[0,1]$ & 0.602 & 2.336 & 1.242\\
\hline
\end{tabular}
	\label{tab:2reg_tab4}
\end{table*}

\end{document}